\documentclass[slac_one]{revtex4}
\usepackage{graphicx}
\usepackage{fancyhdr}
\begin{document}

\title{{\small{2005 International Linear Collider Workshop - Stanford,
U.S.A.}}\\ 
\vspace{12pt}
Trilinear Gauge Couplings from $\gamma\gamma\rightarrow W^{+}W^{-}$} 

%

\author{K. M\"onig, J. Sekaric}
\affiliation{DESY, Platanenallee 6, 15738 Zeuthen, Germany}
%

\begin{abstract}
  If there is no the Standard Model Higgs boson, the interaction among the
  gauge bosons becomes strong at high energies ($\sim 1 \, {\rm TeV}$). The
  effects of strong electroweak symmetry breaking could manifest themselves
  indirectly through the vertices as anomalous gauge boson couplings before
  they give rise to new physical states like resonances. Here a study of the
  measurement of trilinear gauge couplings $\kappa_{\gamma}$ and
  $\lambda_{\gamma}$ is presented looking at the hadronic decay channel of the
  WW boson pair at an $\gamma\gamma$ - collider. A sensitivity of ${\cal
    O}(10^{-3}-10^{-4})$ can be reached depending on the coupling under
  consideration and on the initial polarisation state.

\end{abstract}

\maketitle

\thispagestyle{fancy}


\section{INTRODUCTION} 
Deviations of the triple gauge boson couplings (TGCs) from their values
predicted by the Standard Model (SM) are a possible indication for new physics
(NP) beyond the SM. If no light Higgs boson exists the mechanism responsible
for the restoring the unitarity could well be the strong electroweak symmetry
breaking (SEWSB) mechanism \cite{strong}. As a consequence, at energies below
NP cut-off scale $\Lambda_{NP}$\footnote{$\Lambda_{NP}\sim 4\pi v\approx 3 \,
  {\rm TeV}$} the effects of NP are reflected in the TGC's values leading to
their deviations $\Delta\kappa_{\gamma}$ and $\Delta\lambda_{\gamma}$ from the
SM predictions. Since these deviations decrease as $\Lambda_{NP}$ increases,
their observation requires a very precise measurements, more precise than
those at LEP and Tevatron. With a high event statistics at a $\gamma\gamma$
collider option at the International Linear Collider (ILC) it is possible to
reach a high precision of the TGC measurements.
\par
Anomalous TGCs affect both the total production cross-section and the shape of
the differential cross-section as a function of the W production angle. As a
consequence, distributions of W decay products are changed also. Thus, the
information about TGCs can be extracted from the angular distributions of the
reconstructed W boson. In $\gamma\gamma$ collisions the TGCs contribute
through \textit{t}-channel W-exchange.
\par
\begin{figure*}[t]
\centering
\includegraphics[width=70mm]{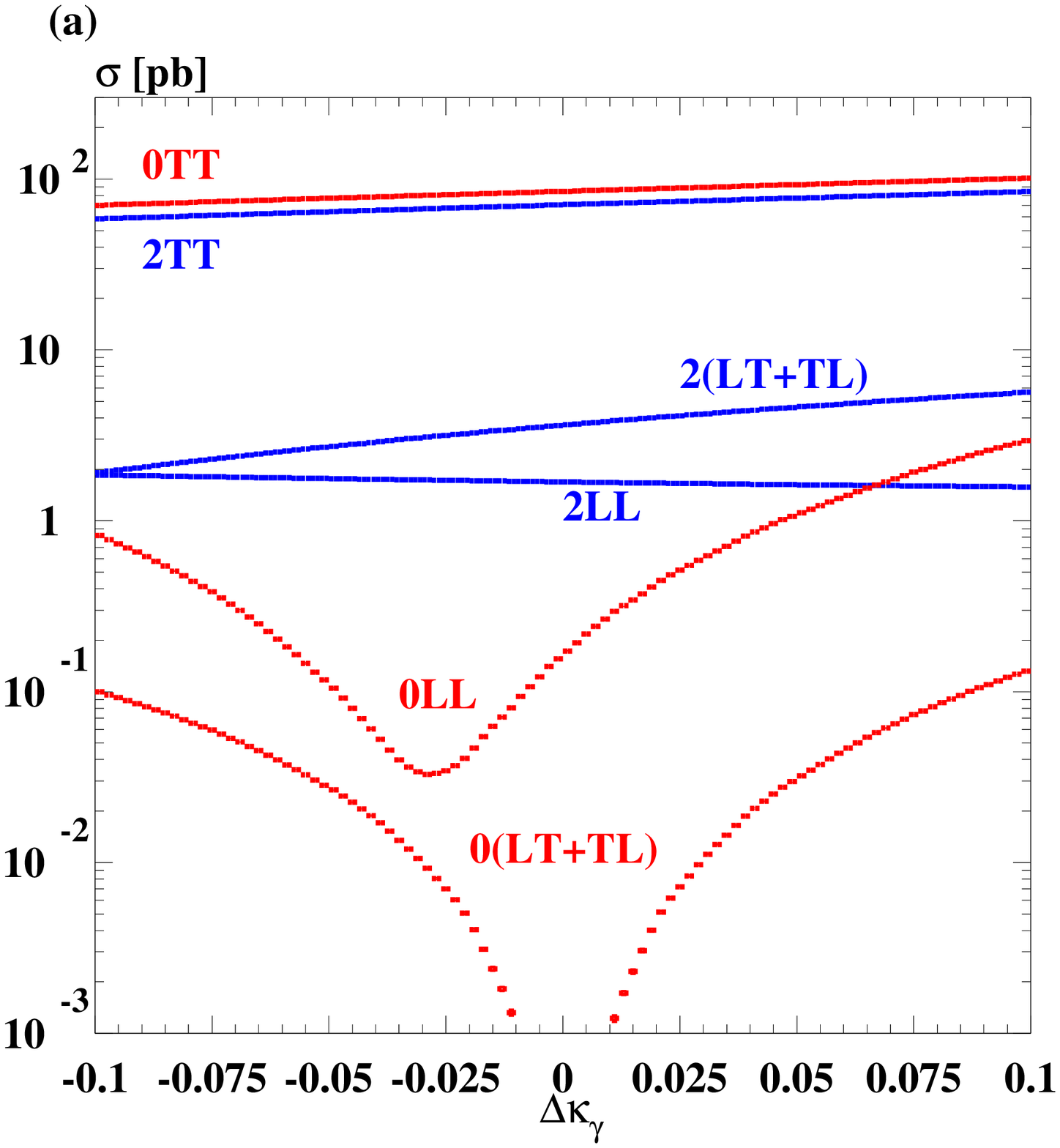}
\includegraphics[width=70mm]{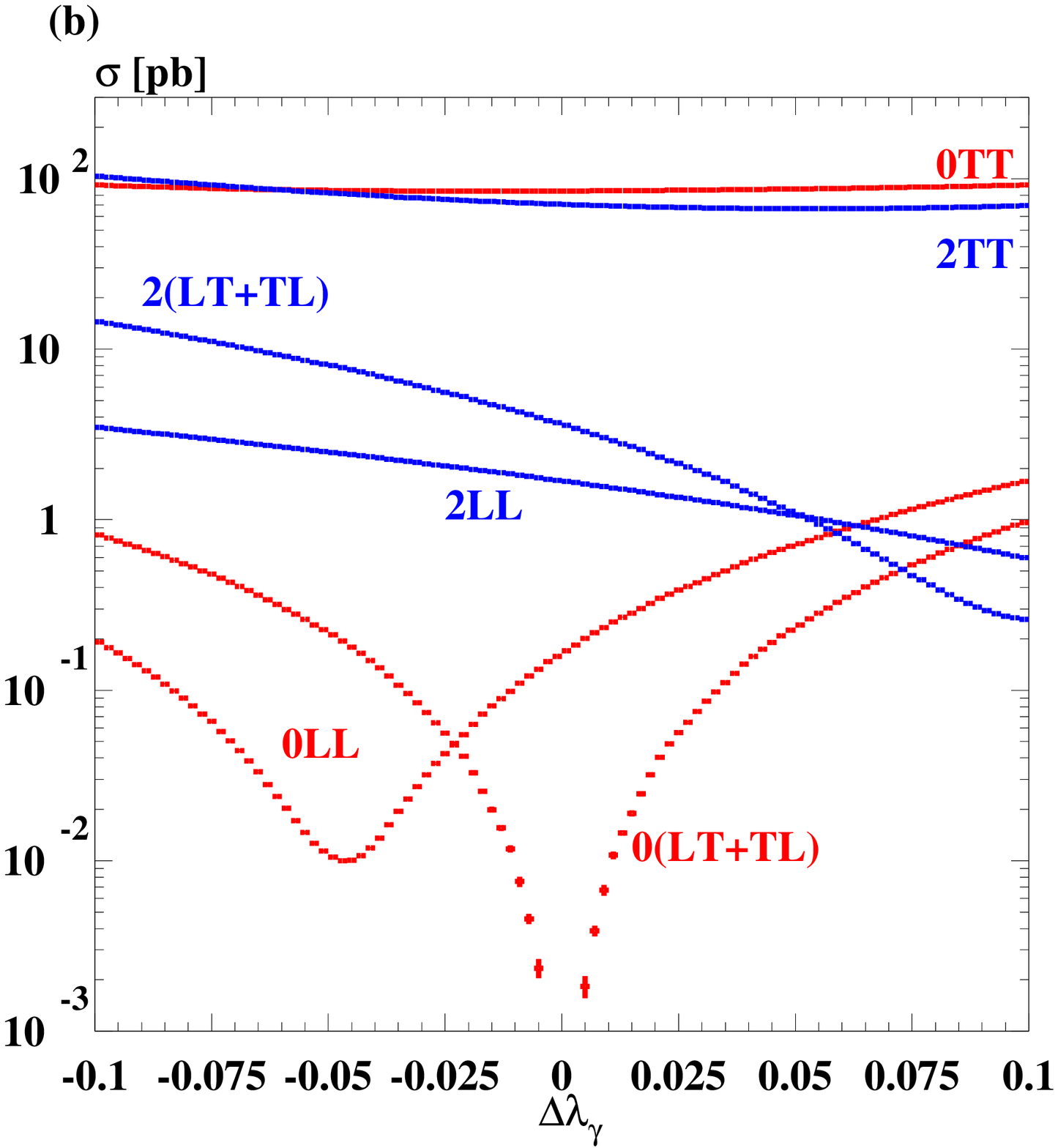}
\caption{Contribution of different $WW$ helicity states for $J_{Z}=0$ and
  $|J_{Z}|=2$ states in the presence of anomalous couplings (\textit{a}):
  ${\kappa}_{\gamma}$ and (\textit{b}): ${\lambda}_{\gamma}$ at
  $\sqrt{s_{\gamma\gamma}}=400$ GeV, assuming fully polarised photon
  beams. The deviations are denoted as $\Delta\kappa_{\gamma}$ and
  $\Delta\lambda_{\gamma}$. The initial photon state $J_{Z}=0$ is denoted with
  ``0'' in front of helicity labelling ($LL,TT,(LT+TL)$) while the state
  $|J_{Z}|=2$ is denoted with ``2''. TT=($\pm\pm$) for $J_{Z}=0$ and
  TT=($\pm\pm$)+($\pm\mp$) for $|J_{Z}|=2$. LT+TL=($\pm 0$)+($0\pm$) and
  LL=($00$).} 
\label{fig:total_gg}
\end{figure*}
\begin{figure*}[t]
\centering
\includegraphics[width=60mm]{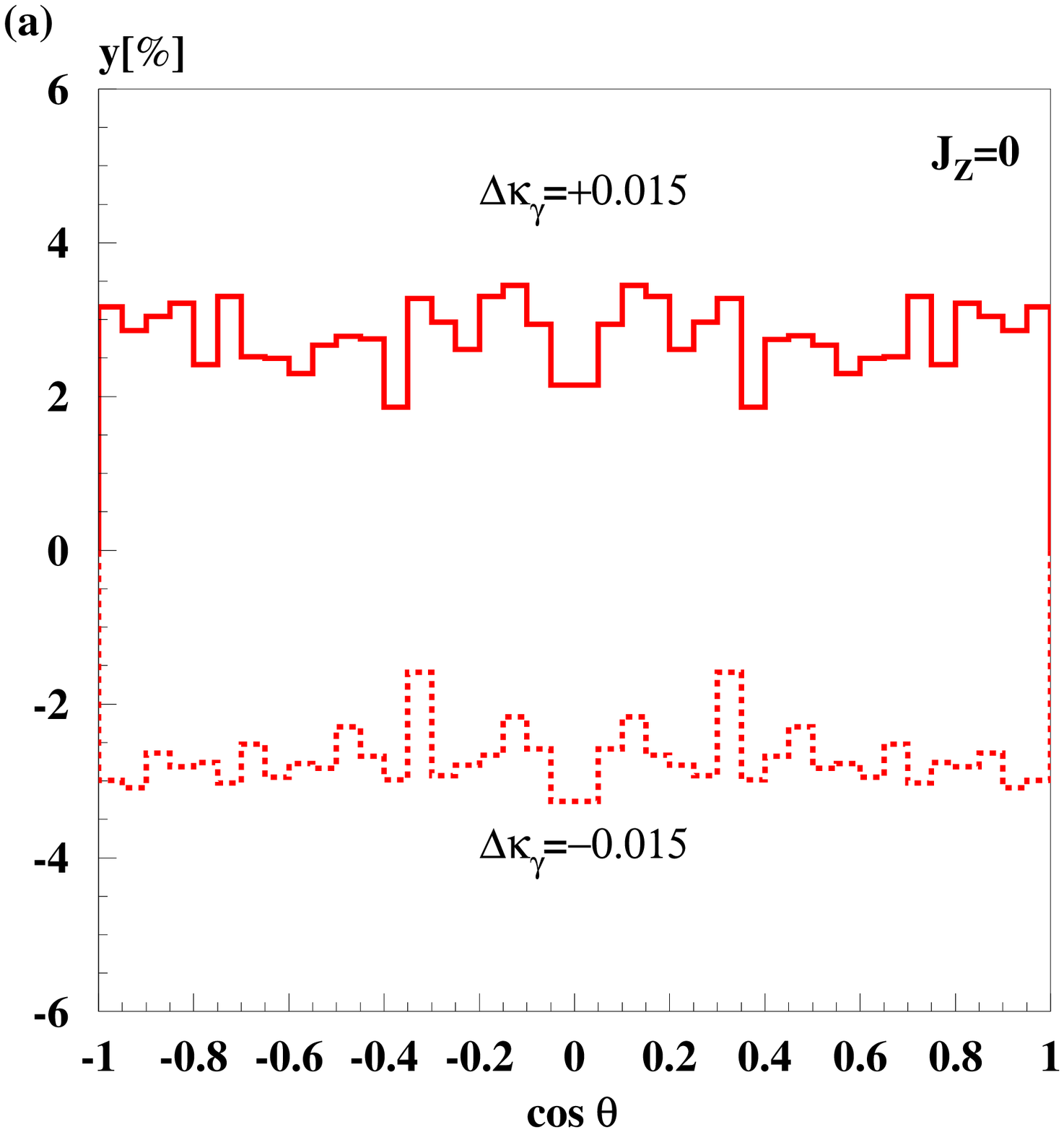}
\includegraphics[width=60mm]{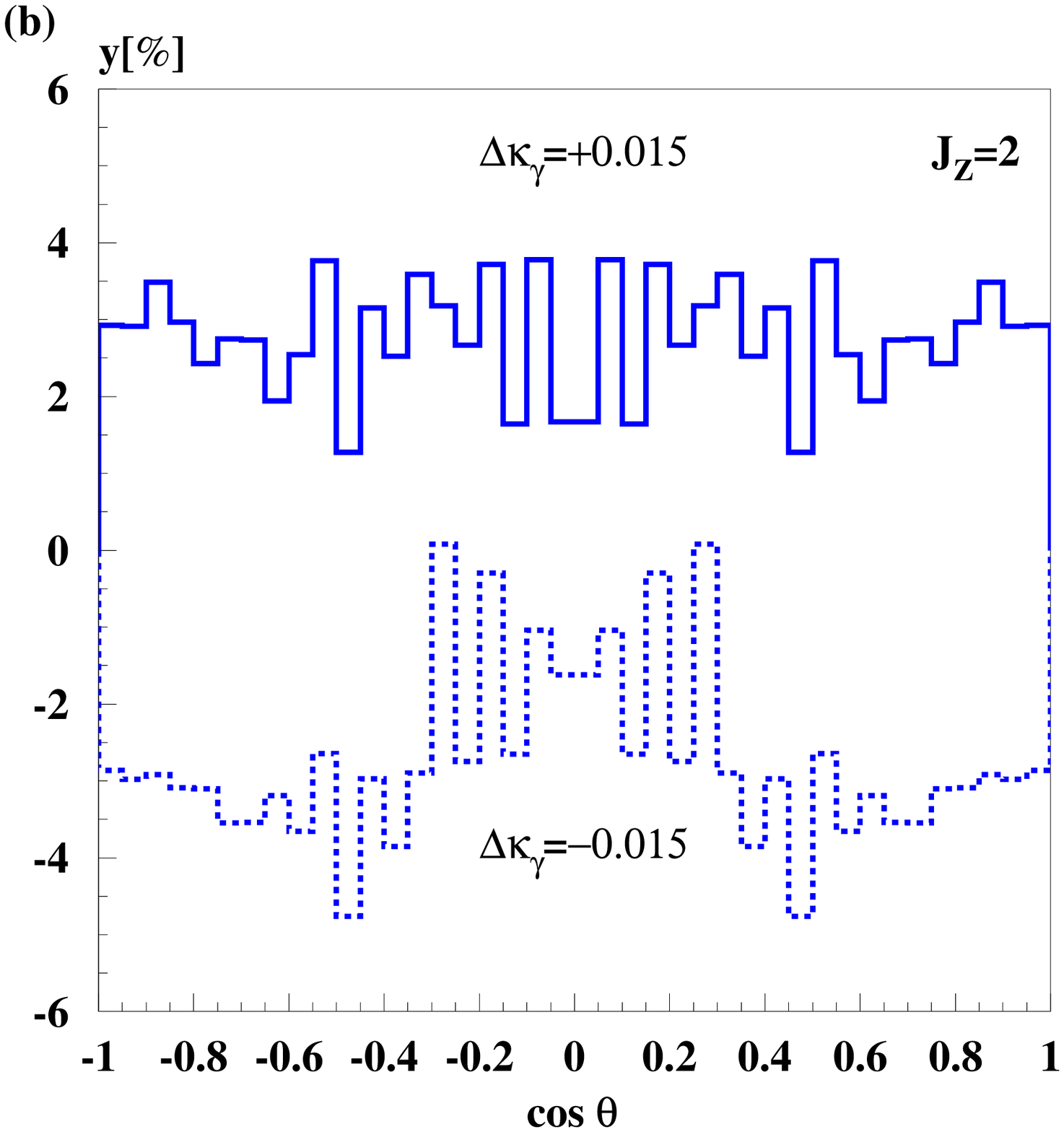}
\includegraphics[width=60mm]{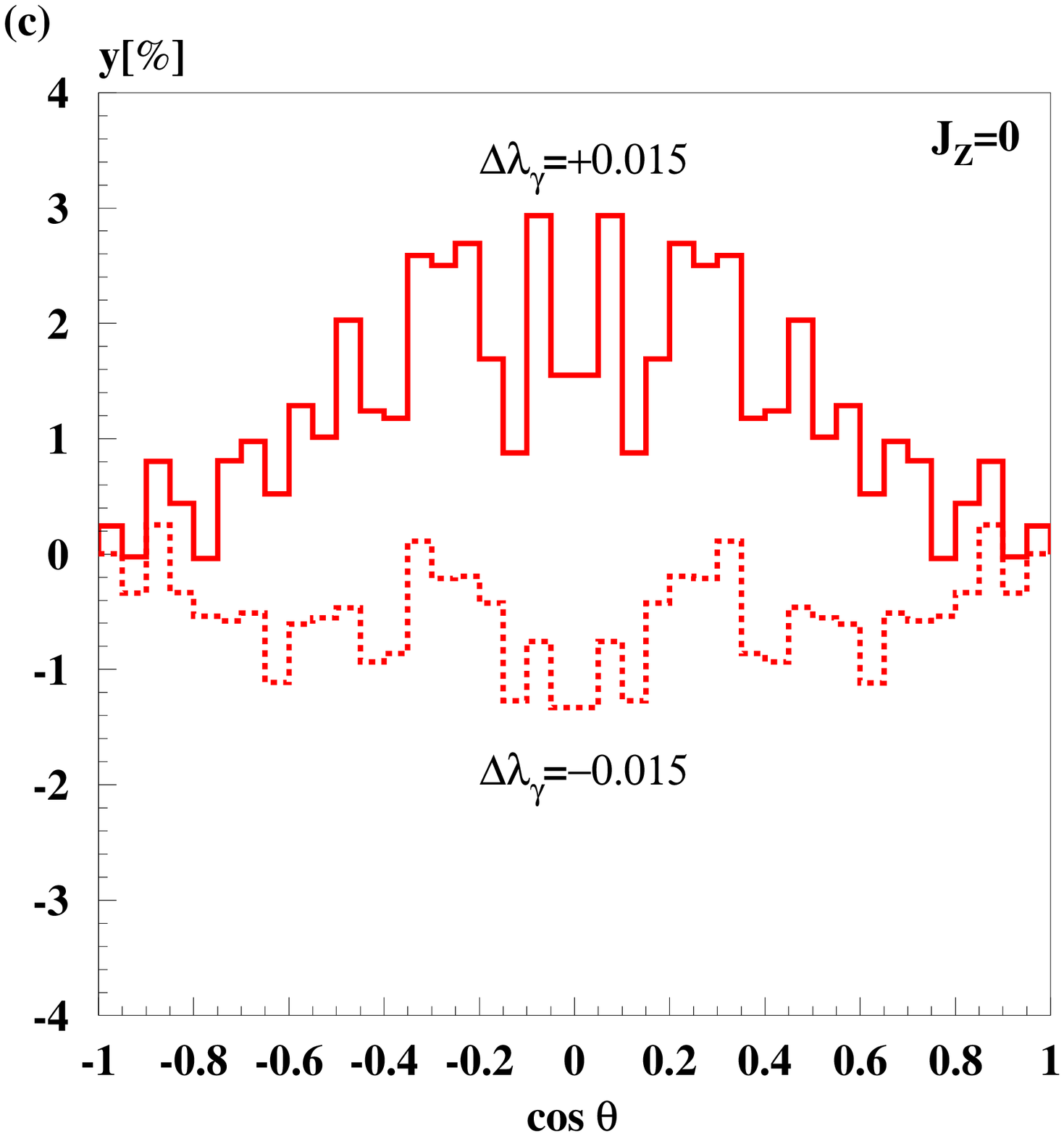}
\includegraphics[width=60mm]{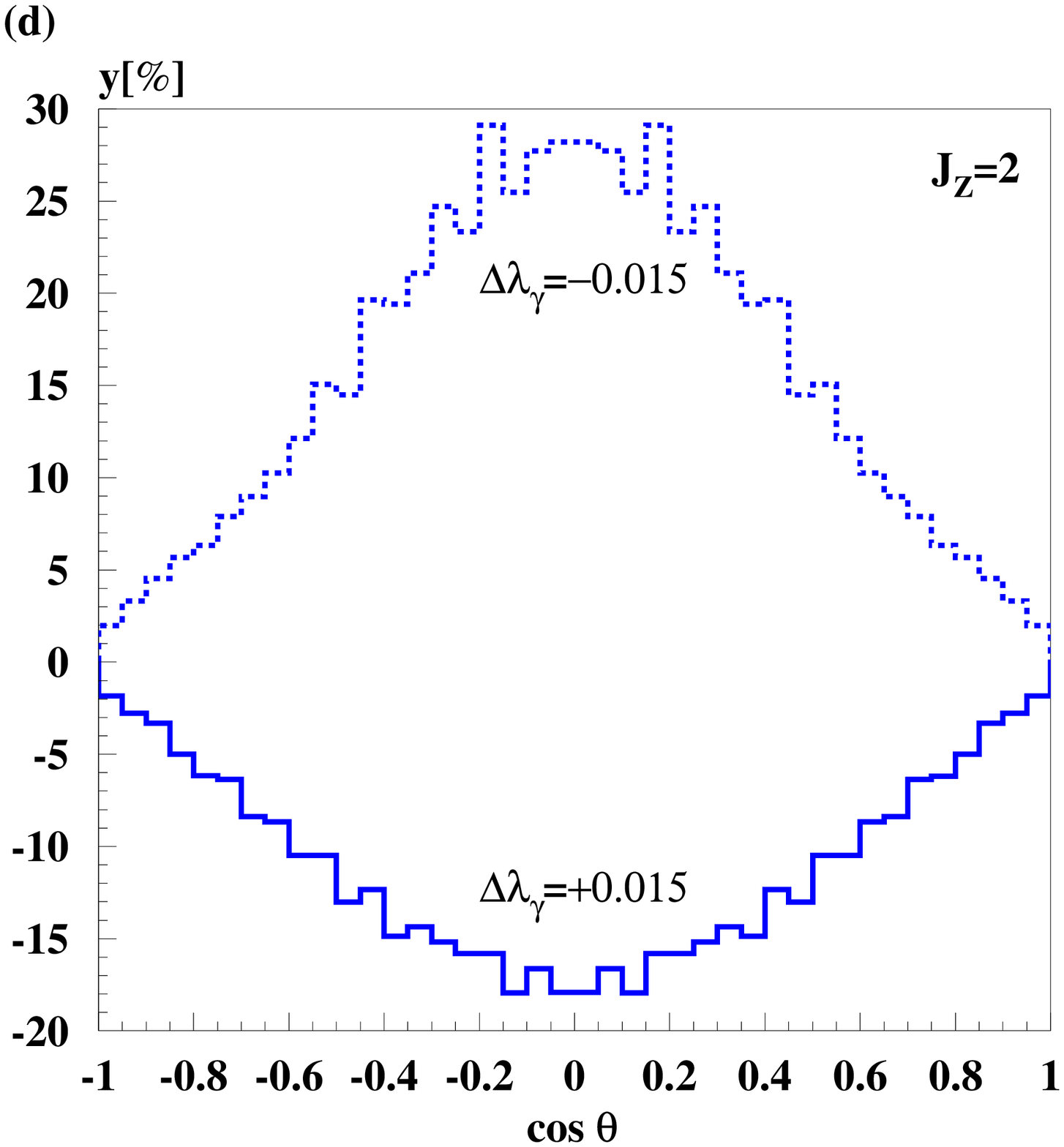}
\caption{Relative deviations of differential cross-section from the SM
  predictions in presence of anomalous coupling (\textit{a}):
  $\kappa_{\gamma}=\pm 1.015$ in the $J_{Z}=0$ state
  (${\Delta}{\lambda}_{\gamma}=0$), (\textit{b}): $\kappa_{\gamma}=\pm 1.015$
  in the $|J_{Z}|=2$ state (${\Delta}{\lambda}_{\gamma}=0$), (\textit{c}):
  $\lambda_{\gamma}=\pm 0.015$ in the $J_{Z}=0$ state
  (${\Delta}{\kappa}_{\gamma}=0$) and (\textit{d}): $\lambda_{\gamma}=\pm
  0.015$ in the $|J_{Z}|=2$ state (${\Delta}{\kappa}_{\gamma}=0$), at
  $\sqrt{s_{\gamma\gamma}}=400$ GeV, assuming fully polarised photon beams.
  Solid lines correspond to
  ${\Delta}{\kappa}_{\gamma},{\Delta}{\lambda}_{\gamma}=+0.015$ and dotted
  lines correspond to
  ${\Delta}{\kappa}_{\gamma},{\Delta}{\lambda}_{\gamma}=-0.015$. All WW
  helicity combinations are included.
  $y[\%]=\frac{\left[d\sigma_{TOT}^{AC}-d\sigma_{TOT}^{SM}\right]}{d\sigma_{TOT}^{SM}}$.}
\label{fig:differ_gg}
\end{figure*}

In this study the expected sensitivity for a measurement of the couplings
$\kappa_{\gamma}$ and $\lambda_{\gamma}$ in $\gamma\gamma\rightarrow
W^{+}W^{-}\rightarrow 4$jets at ${\sqrt{s_{ee}}=500}\,{\rm GeV}$
($\sqrt{s_{\gamma\gamma}} \le 400\,{\rm GeV}$) is investigated. There are two
possible initial $\gamma\gamma$ helicity states depending on the photon
handedness, denoted as $J_{Z}=0$ (if two photons have the same helicities) and
$|J_{Z}|=2$ (if two photons have the opposite helicities). Total and
differential cross-sections distributions as a function of the anomalous TGCs
($\Delta\kappa_{\gamma},\Delta\lambda_{\gamma}\ne 0$), simulated with the
tree-level Monte Carlo (MC) generator WHIZARD \cite{whizard}, for all possible
initial and final state helicity combinations are shown in
Figures~\ref{fig:total_gg} and \ref{fig:differ_gg}.

\section{SIGNAL AND BACKGROUND SIMULATION}
As a beam simulation CIRCE2 \cite{circe2} is used to describe realistic beam
spectra for $\gamma\gamma$-colliders. The response of a detector has been
simulated with SIMDET V4 \cite{simdet4}, a parametric Monte Carlo for the
TESLA $e^{+}e^{-}$-detector. It includes a tracking and calorimeter simulation
and a reconstruction of energy-flow-objects (EFO)\footnote{Electrons, photons,
  muons, charged and neutral hadrons and unresolved clusters that deposit
  energy in the calorimeters.}. Only the EFOs with a polar angle above
$7^{\circ}$ are taken for the W boson reconstruction, simulating the
acceptance of the photon collider detector as the only difference to the
$e^{+}e^{-}$-detector \cite{ggparis}. The signal and background events are
studied on a sample of events generated with WHIZARD and overlayed with low
energy ${\gamma\gamma}{\rightarrow}{\rm hadrons}$ events (\textit{pileup})
\cite{thesis}. The corresponding number of added pileup events per bunch
crossing is 1.8 \cite{schulte}. The informations about the neutral particles
(\textit{neutrals}) from calorimeter and charged tracks (\textit{tracks}) from
tracking detector are used to reconstruct the signal and background events.
The potential background for both initial $J_{Z}$ states are
$\gamma\gamma\rightarrow {q}{\bar{q}}$ events that can mimic the signal with
four jets when gluons are radiated in the final state. The QCD corrections
to the $q\bar{q}$-pair Born level production cross-section are different for
the two $J_{Z}$ states: in the $|J_{Z}|=2$ state the corrected cross-section
is $\sigma_{2}^{QCD}\sim\sigma_{2}^{Born}(1+k\alpha_{s}/\pi)$ with $k$ being
of ${\cal O}$(1), resulting in a Born cross-section correction of 4-5$\%$.
In this study this correction is not taken into account. In the $J_{Z}=0$
state, the suppression factor $(m_{f}^{2}/s)$ \cite{fadin} leads to a Born
level cross-section close to zero but the QCD corrections lead to an
enhancement by double-logarithmic terms
$\sim(\alpha_{s}\log^{2}(s/m_{q}^{2}))^{n}$ \cite{dble_log}. To estimate the
corrected cross-section for the $J_{Z}=0$ state the ${\cal O}(\alpha_{s}^{2})$
diagrams are taken into account i.e. the diagrams contributing to
$\gamma\gamma \rightarrow q\bar{q}gg$ and $\gamma\gamma\rightarrow
q\bar{q}(g\rightarrow)q\bar{q}$. The $y_{cut}$ cut parameter
($(p_{a}+p_{b})^{2}>sy_{cut}$; $a,b=q,\bar{q},g/q,g/\bar{q}$) for a variable
centre-of-mass energy $s$ is defined by generating only events with the
invariant masses of each parton pair above 30 GeV resulting in an emission
of hard gluons. The signal events for both $J_{Z}$ states and background
events for the $|J_{Z}|=2$ state are generated with O'Mega matrix element
generator \cite{omega} taking into account only the lowest order Feynman
diagrams. The QCD correction for the $q\bar{q}$ pair production in the
$J_{Z}=0$ state is estimated generating the background events with MadGraph
\cite{madgraph}.

\subsection{Energy Flow and Event Selection}
\begin{figure*}[t]
\centering
\includegraphics[width=60mm]{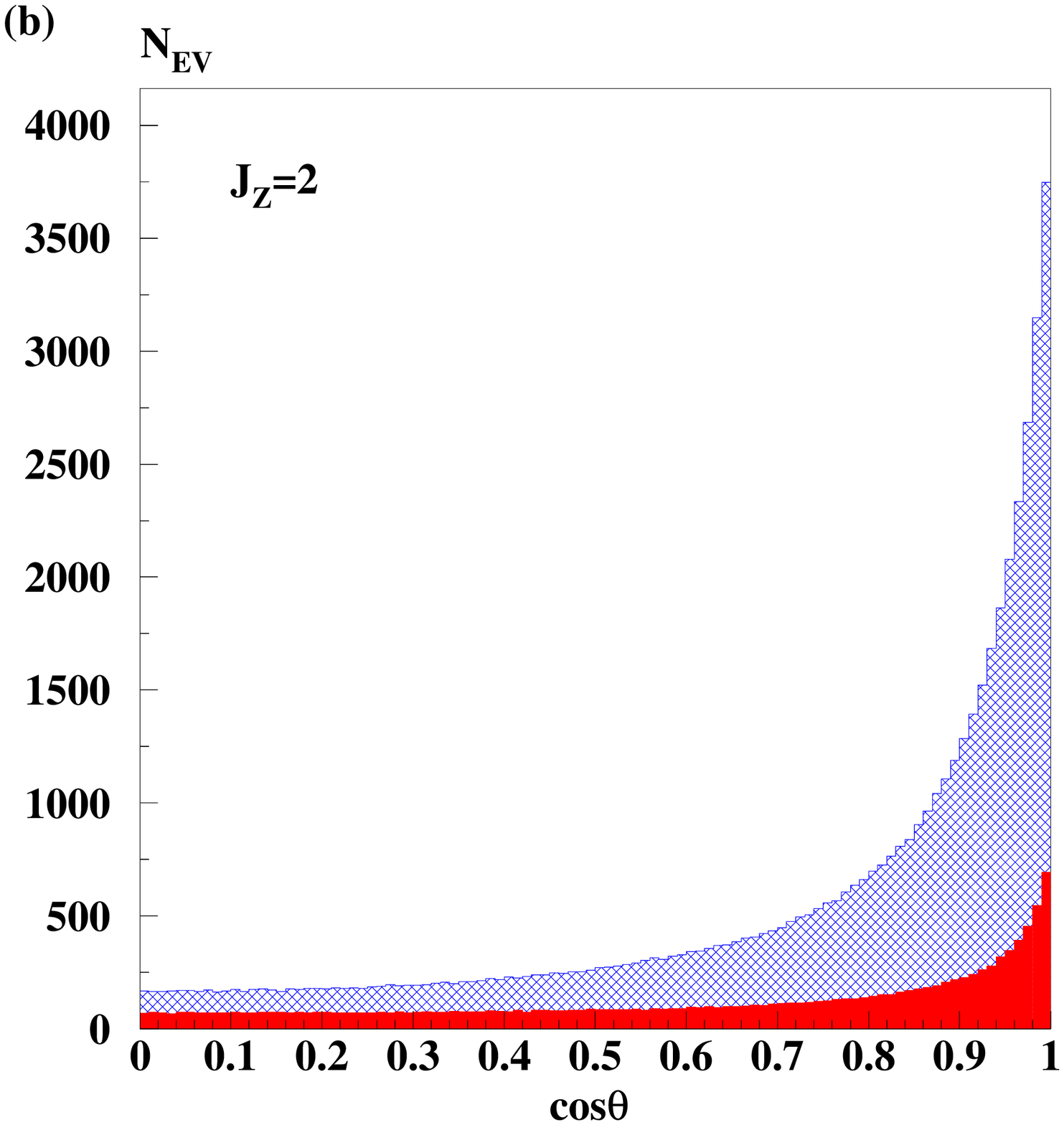}
\includegraphics[width=60mm]{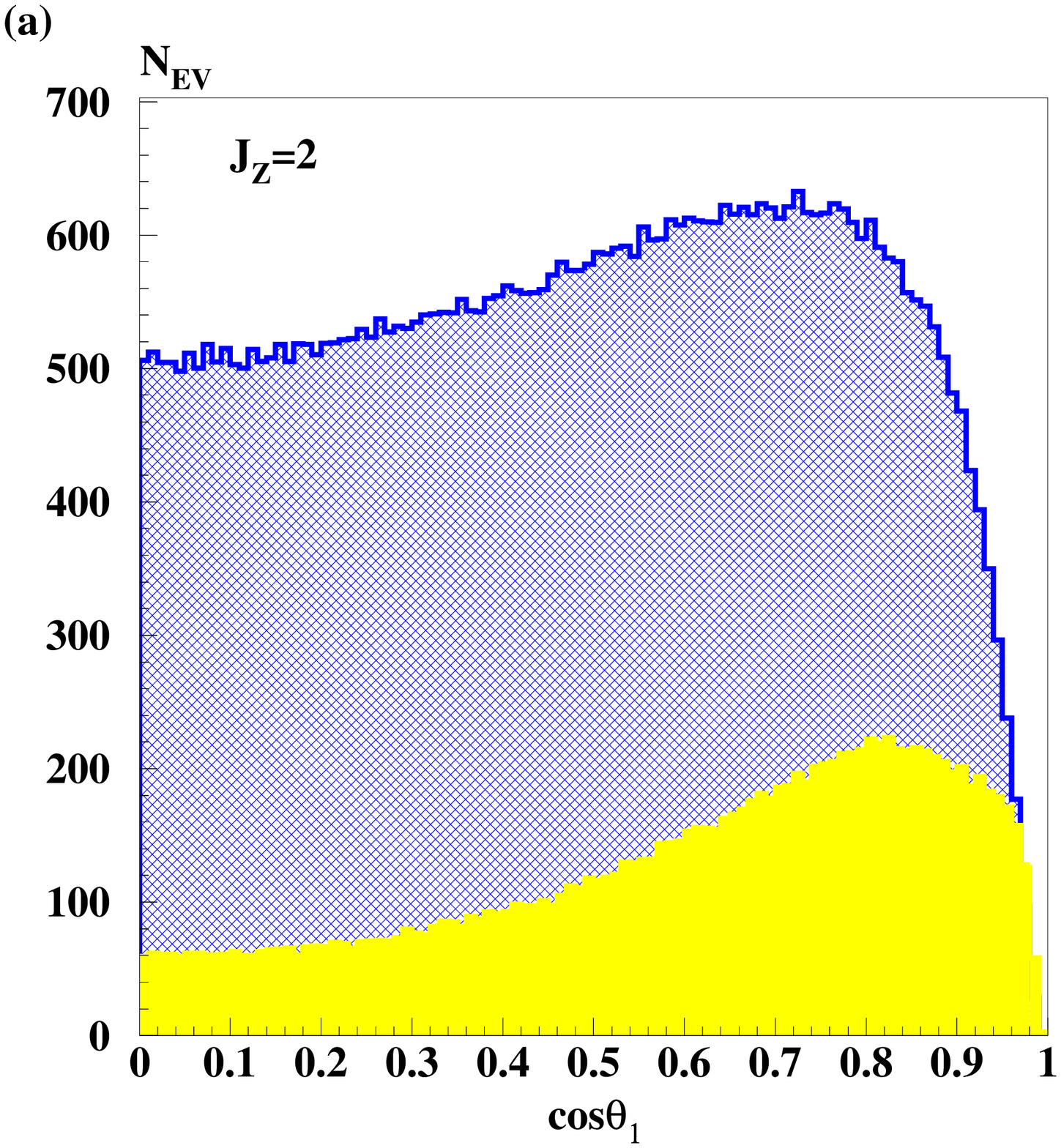}
\includegraphics[width=60mm]{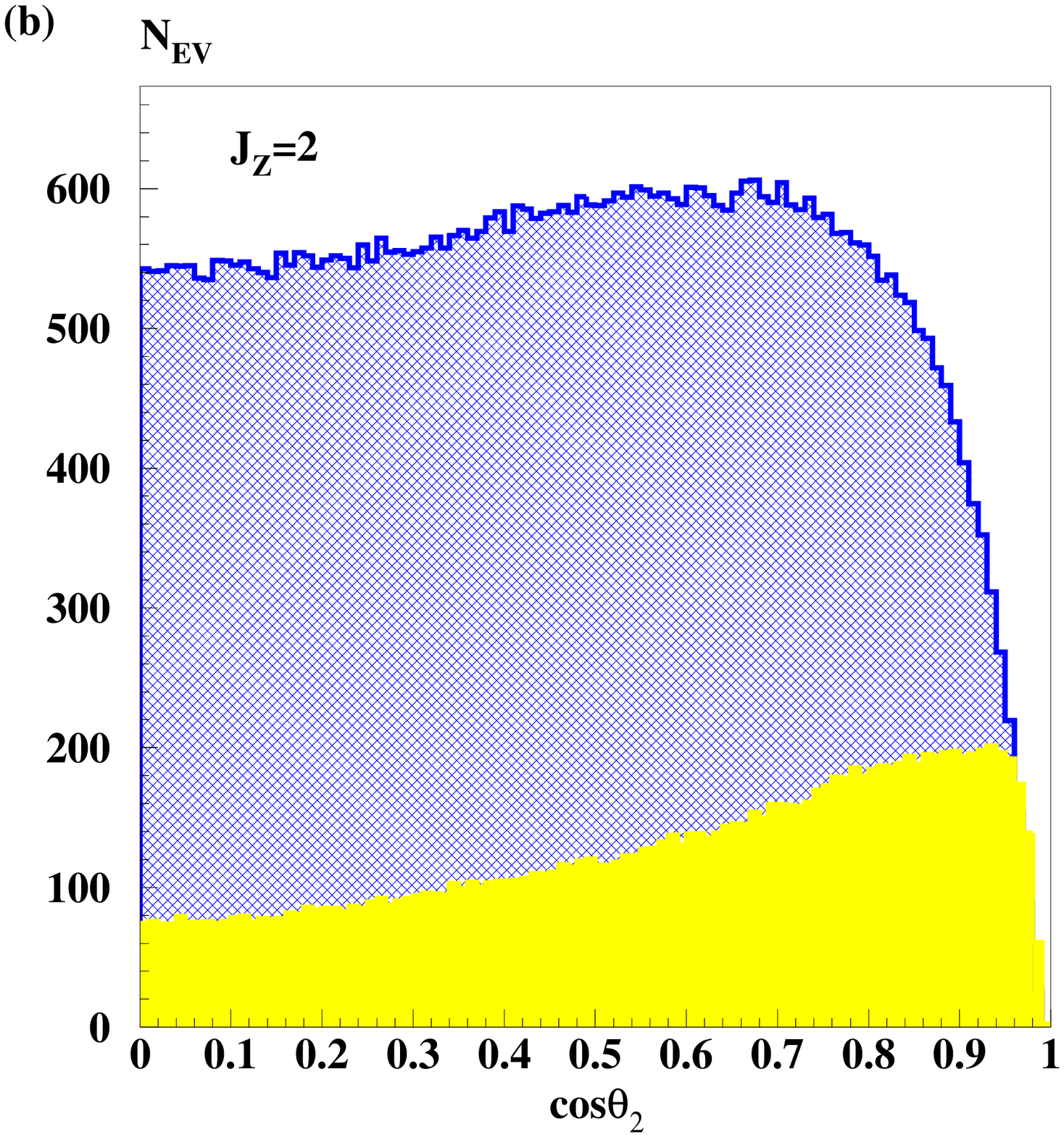}
\includegraphics[width=60mm]{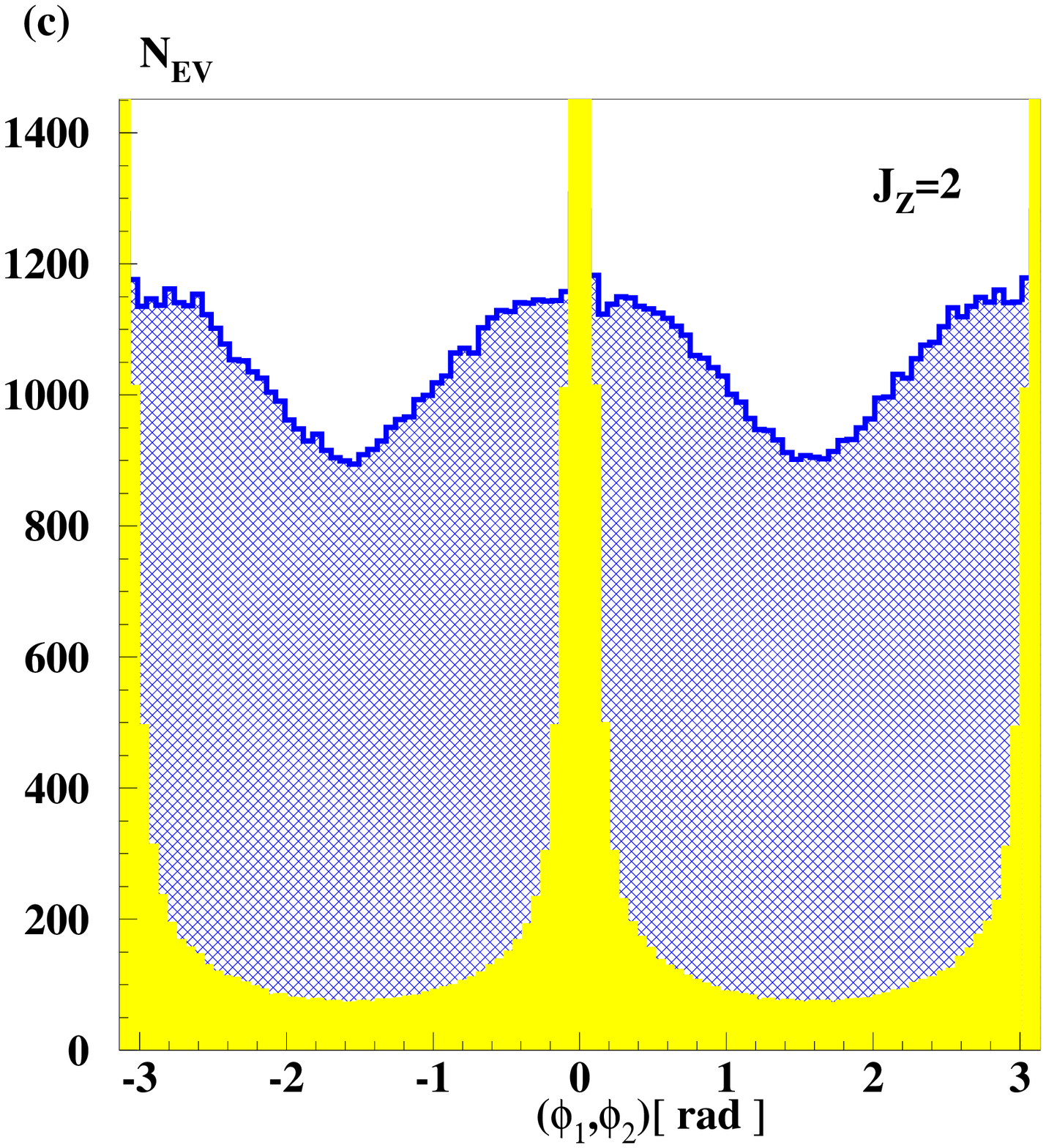}
\caption{Angular distributions of signal (blue) and background (red, yellow)
  events after the detector simulation used in the fit in the $|J_{Z}|=2$
  state over (\textit{a}): the decay angle $\cos\theta_{1}$ of $W_{F}$
  (\textit{b}): the decay angle $\cos\theta_{2}$ of $W_{B}$ (\textit{c}): the
  azimuthal angle ($\phi_{1},\phi_{2}$).} 
\label{fig:final}
\end{figure*}

In order to minimise the pileup contribution to the high energy signal tracks,
the information on the track impact parameters is used in the same way as in
the case of $\gamma e$-collisions \cite{paper} allowing the rejection of $\sim
60\%$ of pileup tracks and $\sim 10-15\%$ of signal tracks. The remaining
tracks are combined into four jets and the events with a number of EFO greater
than 40 and number of charged tracks greater than 20 are accepted only. The two
reconstructed W bosons are denoted as forward ($\cos\theta >0$, $W_{F}$) and
backward ($\cos\theta <0$, $W_{B}$) where $\theta$ is a W boson production
angle in the centre-of-mass system (CMS). The angle between the two jets
belonging to the same W boson, boosted to the CMS, is used as a next selection
criteria - if the angle is within a given range of $40^{\circ} <\theta
<140^{\circ}$, the event is accepted. Further, events with a total mass above
125 GeV and the individual W boson mass of $60\textrm{GeV}< M_{W}
<100\textrm{GeV}$ are accepted. That results in efficiencies of approximately
53\% for signal and less than 2\% for background events i.e. in a
purity of 81\% in both $J_{Z}$ states. The top pair production is estimated
to be negligible. The final angular distributions for the $|J_{Z}|=2$
state\footnote{The similar angular distributions are obtained for the
  $J_{Z}=0$ state.} used for the TGCs error estimation are shown in
Figure~\ref{fig:final}.

\section{FIT METHOD AND ERROR ESTIMATIONS}
For the extraction of the TGSs from the reconstructed kinematical variables
(Fig.~\ref{fig:final}) a binned Likelihood fit is used. A sample of ${2\cdot
  10^{6}}$ SM signal events is generated with WHIZARD and passed trough the
detector simulation. Each event is described reconstructing five kinematical
variables - the W production angle with respect to the ${e^{-}}$ beam
direction $\theta$, the W's polar decay angles $\theta_{1,2}$ (angle of the
fermion with respect to the W flight direction measured in the W rest frame)
and the azimuthal decay angles $\phi_{1,2}$ of the fermion with respect to a
plane defined by W and the beam axis. In hadronic W-decays the up- and
down-type quarks cannot be separated so that only $|cos\theta_{1,2}|$ is
measured. The matrix element calculations from WHIZARD are used to obtain
weights \cite{paper} to reweight the angular distributions as functions of the
anomalous TGCs where ${\Delta}{\kappa}_{\gamma}$ and ${\lambda}_{\gamma}$ are
the free parameters. Six-dimensional (6D) event distributions over
$\cos{\theta}$, $\cos{\theta_{1,2}}$, $\phi_{1,2}$ and centre-of-mass energy
are fitted with MINUIT \cite{min}, minimising the Likelihood function
depending on $\kappa_\gamma$ and $\lambda_\gamma$:
\begin{eqnarray*}
{\cal L} & = & -\sum_{i,j,k,l,m,p}[z\cdot N^{SM}(i,j,k,l,m,p)\cdot\log \left(z\cdot n \cdot N^{{\Delta\kappa}_{\gamma},{\Delta\lambda}_{\gamma}}(i,j,k,l,m,p)\right) \\
& & - z\cdot n \cdot N^{{\Delta\kappa}_{\gamma},{\Delta\lambda}_{\gamma}}(i,j,k,l,m,p)] +\frac{(n-1)^{2}}{2(\Delta L^{2})},
\end{eqnarray*}
where \textit{i,j,k,l} and \textit{m} run over the reconstructed angular
distributions $\cos{\theta},\cos{\theta_{1,2}}$ and $\phi_{1,2}$, \textit{p}
runs over the reconstructed centre-of-mass energy, $N^{SM}(i,j,k,l,m,p)$ is
the ``data'' which corresponds to the SM MC sample,
$N^{{\Delta\kappa}_{\gamma}{\Delta\lambda}_{\gamma}}$$(i$, $j$, $k$, $l$, $m$,
$p)$ (MC sample) is the event distribution weighted by the function
$R({\Delta}{\kappa}_{\gamma}$,${\Delta\lambda}_{\gamma})$ and $\sigma$$(i$,
$j$, $k$, $l$, $m$, $p)=\sqrt{N^{SM}(i,j,k,l,m,p)}$. The factor ${z}$ sets the
number of signal events to the expected one after one year of running of an
$\gamma\gamma$-collider. In case where the background is included in the fit
${z}$ defines the sum of signal and background events and ${n}\cdot
N^{{\Delta\kappa}_{\gamma},{\Delta\lambda}_{\gamma}} \rightarrow [{n} \cdot
N_{signal}^{{\Delta\kappa}_{\gamma},{\Delta\lambda}_{\gamma}}+N_{bck}]$. The
number of background events is normalised to the effective W boson production
cross-section in order to obtain the corresponding number of background events
after one year of running of an $\gamma\gamma$-collider for corresponding
$J_{Z}$ state. It is assumed that the total normalisation (efficiency,
luminosity, electron polarisation) is only known with a relative uncertainty
$\Delta L$. Thus, $n$ is taken as a free parameter in the fit and constrained
to unity with the assumed normalisation uncertainty. Per construction the fit
is bias-free and thus returns always exactly the SM as central values. In the
$|J_{Z}|=2$ state $\Delta L=0.1\%$ is a realistic precision that can be
achieved while for the $J_{Z}=0$ due to the small number of events\footnote{It
  is assumed that the luminosity will be measured counting the events produced
  in $\gamma\gamma\rightarrow l^{+}l^{-}$ where the cross-section is $m^{2}/s$
  suppressed.}, the luminosity is expected to be measured with an error of
$\Delta L=1\%$.
\par
Table \ref{tab:t1} shows the estimated statistical errors we expect for the
different couplings at ${\sqrt{s_{ee}}=500}\,{\rm GeV}$ for a
two-parameter\footnote{A two-parameter fit means that both couplings are
  allowed to vary freely as well as the normalisation \textit{n}.} 6D fit at
detector level including the pileup and background events in both $J_{Z}$
states.

\begin{table}[htb]
\begin{center}
\begin{tabular}{|l||c|c|c||c|c|c|c|c|c|} \hline
\multicolumn{1}{|c||}{1000 fb$^{-1}$} & \multicolumn{3}{|c||}{without pileup} & \multicolumn{3}{c|}{with pileup} & \multicolumn{3}{|c|}{pileup+background} \\ \hline
\multicolumn{1}{|c||}{6D fit} & \multicolumn{3}{|c||}{$J_{Z}=0/|J_{Z}|=2$} & \multicolumn{3}{c|}{$J_{Z}=0/|J_{Z}|=2$} & \multicolumn{3}{|c|}{$J_{Z}=0/|J_{Z}|=2$} \\ \hline \hline
${{\Delta}L}$ & 1$\%$ & 0.1$\%$ & 0 & 1$\%$ & 0.1$\%$ & 0 & 1$\%$ & 0.1$\%$ & 0 \\ \hline\hline
${\Delta}{\kappa}_{\gamma}{\cdot}10^{-4}$ & 19.9/29.9 & 5.5/6.2 & 2.6/3.7 & 26.9/37.4 & 5.8/6.8 & 3.0/4.6 & 27.8/37.8 & 5.9/7.0 & 3.1/4.8 \\ \hline
${\Delta}{\lambda}_{\gamma}{\cdot}10^{-4}$ & 3.7/3.1 & 3.7/3.1 & 3.7/3.1 & 5.4/4.6 & 5.2/4.6 & 5.2/4.6 & 5.7/4.8 & 5.6/4.8 & 5.6/4.8 \\ \hline
\end{tabular}
\end{center}
\caption{Estimated statistical errors for ${\kappa}_{\gamma}$ and
  ${\lambda}_{\gamma}$ from the 6D fit at detector level for both $J_{Z}$
  states in $\gamma\gamma$ collisions at $\sqrt{s_{ee}}=500$ GeV, without
  pileup, with pileup and with background events.} 
\label{tab:t1}
\end{table}

In Table \ref{tab:t2} the results for  
$\sqrt{s_{\gamma \gamma}}=400\,{\rm GeV}$ and 
$\sqrt{s_{\gamma \gamma}}=400\,{\rm GeV}$ are compared using a fixed photon
energy. 

\begin{table}[htb]
\begin{center}
\begin{tabular}{|l||c|c|c||c|c|c||c|c|c||c|c|c|} \hline
\multicolumn{1}{|c||}{110 fb$^{-1}$} & \multicolumn{6}{|c||}{$\sqrt{s_{\gamma\gamma}}=400$ GeV} & \multicolumn{6}{|c|}{$\sqrt{s_{\gamma\gamma}}=800$ GeV} \\ \hline
\multicolumn{1}{|c||}{5D fit} & \multicolumn{3}{|c||}{$J_{Z}=0$} & \multicolumn{3}{|c||}{$|J_{Z}|=2$} & \multicolumn{3}{|c||}{$J_{Z}=0$} & \multicolumn{3}{|c|}{$|J_{Z}|=2$} \\ \hline
${{\Delta}L}$ & 1$\%$ & 0.1$\%$ & 0 & 1$\%$ & 0.1$\%$ & 0 & 1$\%$ & 0.1$\%$ & 0 & 1$\%$ & 0.1$\%$ & 0 \\ \hline\hline
${\Delta}{\kappa}_{\gamma}{\cdot}10^{-4}$ & 14.4 & 5.4 & 2.6 & 20.1 & 6.2 & 3.8 & 7.2 & 4.5 & 2.4 & 8.1 & 4.6 & 2.6 \\ \hline
${\Delta}{\lambda}_{\gamma}{\cdot}10^{-4}$ & 3.0 & 3.0 & 3.0 & 1.6 & 1.6 & 1.6  & 1.3 & 1.3 & 1.3 & 0.63 & 0.58 & 0.56 \\ \hline
\end{tabular}
\end{center}
\caption{Estimated statistical errors for ${\kappa}_{\gamma}$ and
  ${\lambda}_{\gamma}$ from the five-dimensional (5D) two-parameter fit at
  generator level for the $J_{Z}=0$ and $|J_{Z}|=2$ at $\gamma\gamma$
  collisions at $\sqrt{s_{\gamma\gamma}}=400$ and 800 GeV. The number of
  events for both $J_{Z}=0$ states is normalised to the expected one with
  integrated luminosity of 110 fb$^{-1}$ in the high energy peak.} 
\label{tab:t2}
\end{table}
\par
The comparison of $\kappa_{\gamma}$ and $\lambda_{\gamma}$ obtained from
$e^{+}e^{-},\gamma e$ and $\gamma\gamma$ at $\sqrt{s_{ee}}=500$ GeV is shown
in Table \ref{tab:t3} (left side). The right side of Table \ref{tab:t3} shows
the comparison at $\sqrt{s_{e^{+}e^{-},\gamma\gamma}}=800$ GeV for the two
types of collider. The sensitivities to $\kappa_{\gamma}$ and
$\lambda_{\gamma}$ in $\gamma\gamma\rightarrow W^{+}W^{-}$ at
$\sqrt{s_{\gamma\gamma}}=800$ GeV, including the variable energy spectrum,
background and pileup events are approximated scaling the estimated
sensitivities at generator level (Table \ref{tab:t2}) by a factor obtained
for $\sqrt{s_{ee}}=500$ GeV. The sensitivities at an $e^{+}e^{-}$-collider are
estimated at generator level.
\begin{table}[htb]
\begin{center}
\begin{tabular}{|c||c||c|c||c||c||c|c||c|} \hline
& \multicolumn{4}{|c||}{$\sqrt{s_{ee}}=500$ GeV} & & \multicolumn{3}{|c|}{$\sqrt{s_{\gamma\gamma,e^{+}e^{-}}}=800$ GeV} \\ \hline\hline
LEFT & {$\gamma e$} & \multicolumn{2}{|c||}{$\gamma\gamma$} & $e^{+}e^{-}$ & RIGHT & \multicolumn{2}{|c||}{$\gamma\gamma$} & $e^{+}e^{-}$ \\ \hline\hline
Mode & Real/Parasitic $|J_{Z}|=3/2$ & $|J_{Z}|=2$ & $J_{Z}=0$ & $|J_{Z}|=1$ & Mode & $|J_{Z}|=2$ & $J_{Z}=0$ & $|J_{Z}|=1$\\ \hline
$\int{\cal L}\Delta t$ & 160 fb$^{-1}$/230 fb$^{-1}$ & \multicolumn{2}{|c||}{1000 fb$^{-1}$} & 500 fb$^{-1}$ & $\int{\cal L}\Delta t$ & \multicolumn{3}{|c|} {1000 fb$^{-1}$} \\ \hline
${{\Delta}L}$ & \multicolumn{1}{|c||}{0.1$\%$} & 0.1$\%$ & 1$\%$ & - & ${{\Delta}L}$ & 0.1$\%$ & 1$\%$ & - \\ \hline\hline
${\Delta}{\kappa}_{\gamma}{\cdot}10^{-4}$ & 10.0/11.0 & 7.0 & 27.8 & 3.6$^*$ & ${\Delta}{\kappa}_{\gamma}{\cdot}10^{-4}$ & 5.2 & 13.9 & 2.1$^*$ \\ \hline
${\Delta}{\lambda}_{\gamma}{\cdot}10^{-4}$ & 4.9/6.7 & 4.8 & 5.7 & 11.0$^*$ & ${\Delta}{\lambda}_{\gamma}{\cdot}10^{-4}$ & 1.7 & 2.5 & 3.3$^*$ \\ \hline
\end{tabular}
\end{center}
\caption{\textit{Left}: Comparison of the $\kappa_{\gamma}$ and
  $\lambda_{\gamma}$ sensitivities at $\gamma e$-, $\gamma\gamma$- and
  $e^{+}e^{-}$-colliders estimated at $\sqrt{s_{ee}}=500$ GeV using the
  polarised beams. In case of photon colliders, the background and the pileup
  are included. ($^*$) denotes the estimation at the generator
  level. \textit{Right}: Comparison of the $\kappa_{\gamma}$ and
  $\lambda_{\gamma}$ sensitivities at $\gamma\gamma$- and
  $e^{+}e^{-}$-colliders estimated at $\sqrt{s_{e^{+}e^{-},\gamma\gamma}}=800$
  GeV using the polarised beams. ($^*$) denotes the estimation at the
  generator level. The sensitivities at $\gamma\gamma$-collider are scaled for
  the background, pileup and the energy spectrum.} 
\label{tab:t3}
\end{table}

\par
Concerning the systematic errors the influence of the background and the
degree of photon polarisation have been investigated, assuming
${{\Delta}L}=0.1\%$ in the $|J_{Z}|=2$ state and ${{\Delta}L}=1\%$ in the
$J_{Z}=0$ state. In the $J_{Z}=0$ state, the polarisation uncertainty of
$0.0021$ for $\kappa_\gamma$ is to less than the statistical error while in
the $|J_{Z}|=2$ state, the polarisation uncertainty of $0.0018$ for
$\kappa_\gamma$ is less than three times the statistical error. The
uncertainty on $\lambda_\gamma$ in both $J_{Z}$ states is found to be
negligible. In the $|J_{Z}|=2$ state the background cross-section should be
known to better than 0.8\% for $\kappa_\gamma$ and to
better than 4\% for $\lambda_\gamma$ if the corresponding systematic
uncertainty should no be larger than the statistical error.
For $J_{Z}=0$ the requirement is 0.6\% for $\lambda_\gamma$ while there are
basically no restrictions for $\kappa_\gamma$.

\section{CONCLUSIONS}
The estimated sensitivity of the TGCs measurement in both $\gamma\gamma$
initial states at $\sqrt{s_{ee}}=500$ GeV with integrated luminosities of
${\cal L}_{\gamma\gamma}\approx 1000$ fb$^{-1}$ is of order $\approx 7\cdot
10^{-4}$ for $\Delta\kappa_{\gamma}$ and higher than $5\cdot 10^{-4}$ for
$\Delta\lambda_{\gamma}$ in the $|J_{Z}|=2$ state assuming $\Delta L/L\approx
10^{-3}$. The state $J_{Z}=0$ takes into account a larger error on the
luminosity measurement of $\Delta L/L\approx 10^{-2}$ resulting in a
sensitivity to $\kappa_{\gamma}$ higher than $3\cdot 10^{-3}$ and to
$\lambda_{\gamma}$ higher than $6\cdot 10^{-4}$. While $\kappa_\gamma$ can be
measured somewhat better in $e^+ e^-$, the $\gamma\gamma$-collider provides a
higher accuracy for a $\lambda_\gamma$ measurement compared to the $e^+ e^-$-
and $\gamma e$-colliders.


\begin{thebibliography}{9}   
  
\bibitem{strong} 
M. Chanowitz, M. Golden and H. Georgi, Phys. Rev. D36 (1987)
  1490; M.J.G. Veltman and F.J. Ynddurain, Nucl. Phys. B325 (1989) 1.
\bibitem{whizard} 
W.\,Kilian, ``WHIZARD 1.24 A generic Monte Carlo integration
  and event generation package for multi-particle processes'', LC-TOOL
  2001-039 (revised) (2001).

\bibitem{circe2} 
T. Ohl, ``Circe Version 2.0: Beam Spectra for Simulating
  Linear Collider and Photon Collider Physics'',
  http://heplix.ikp.physik.tu-darmstadt.de/pub/ohl/circe2.
  
\bibitem{simdet4} 
M.Pohl, H.J.Schreiber, ``SIMDET-Version 4 A parametric Monte
  Carlo for a TESLA Detector'', DESY 02-061, May 2002.

\bibitem{ggparis}
K.\,M\"onig, ``A Photon Collider at TESLA'', LC-DET-2004-014 (2004).

\bibitem{thesis} 
D. Schulte, ``Study of Electromagnetic and Hadronic
  Background in the Interaction Region of the TESLA Collider'', Thesis, April
  1997.

\bibitem{schulte}
D. Schulte, private communication.

\bibitem{fadin} 
V.S.\,Fadin, V.A.\,Khoze and A.D.\,Martin, 
Phys. Rev. D56 (1997) 484-503.

\bibitem{dble_log} 
M.\,Melles and W.J.\,Stirling, 
Phys. Rev. D59 (1999) 094009.

\bibitem{omega}
M.\,Moretti, T.\,Ohl and J.\,Reuter, LC-TOOL-2001-040 (2001).

\bibitem{madgraph}
T.\,Stelzer and W.F.\,Long, Comput. Phys. Commun.81 (1994) 357.

\bibitem{paper} 
K.\,M\"onig and J.\,Sekaric, 
Eur.\,Phys.\,J.C.\,38 (2005) 427-436.
\bibitem{min} F.James, MINUIT Function Minimization and Error Analysis,
  Version 94.1, CERN Program Library Long Writeup D506.

\end{thebibliography}
\end{document}